\documentclass[]{spie}  %>>> use for US letter paper
\usepackage{amsmath,amsfonts,amssymb}
\usepackage{graphicx}
\usepackage[colorlinks=true, allcolors=blue]{hyperref}
\graphicspath{ {images/} }
\usepackage[skip=10pt]{caption} % example skip set to 2pt
\usepackage{subcaption}
\usepackage[export]{adjustbox}
\usepackage{pdfpages}
\usepackage{placeins}
\usepackage{xcolor}
\usepackage[utf8]{inputenc}

\usepackage{pdflscape}

\usepackage{float}

\title{Dark zone maintenance results for segmented aperture wavefront error drift in a high contrast space coronagraph}

\author[a]{Susan F. Redmond}
\author[b]{Laurent Pueyo}
\author[c]{Leonid Pogorelyuk}
\author[b]{Emiel Por}
\author[b]{James Noss}
\author[b]{Keira Brooks}
\author[d,e]{Iva Laginja}
\author[b,f]{Scott D. Will}
\author[b]{Marshall D. Perrin}
\author[b]{Remi Soummer}
\author[a,g]{N. Jeremy Kasdin}

\affil[a]{Princeton University, Olden Street, Princeton, NJ 08544, USA}
\affil[b]{Space Telescope Science Institute, 3700 San Martin Drive, Baltimore, MD 21218, USA}
\affil[c]{Massachusetts Institute of Technology,77 Massachusetts Ave, Cambridge, MA 02139, USA}
\affil[d]{DOTA, ONERA, Université Paris Saclay, F-92322 Châtillon, France}
\affil[e]{Aix Marseille Université, CNRS, LAM (Laboratoire d’Astrophysique de Marseille) UMR 7326, 13388 Marseille, France}
\affil[f]{University of Rochester, 500 Joseph C. Wilson Blvd., Rochester, NY 14627, USA}
\affil[g]{University of San Francisco, 2130 Fulton St, San Francisco, CA 94117, USA}

\authorinfo{Further author information: (Send correspondence to S.F.R.)\\S.F.R.: E-mail: sredmond@princeton.edu}

\begin{document} 
\maketitle

\begin{abstract}
Due to the limited number of photons, directly imaging planets requires long integration times with a coronagraphic instrument. The wavefront must be stable on the same time scale, which is often difficult in space due to thermal variations and other mechanical instabilities. In this paper, we discuss the implications on future space mission observing conditions of our recent laboratory demonstration of a dark zone maintenance (DZM) algorithm.  The experiments are performed on the High-contrast imager for Complex Aperture Telescopes (HiCAT) at the Space Telescope Science Institute (STScI).  The testbed contains a segmented aperture, a pair of continuous deformable mirrors (DMs), and a lyot coronagraph. The segmented aperture injects high order wavefront aberration drifts into the system which are then corrected by the DMs downstream via the DZM algorithm.  We investigate various drift modes including segmented aperture drift, all three DMs drift simultaneously, and drift correction at multiple wavelengths. 
\end{abstract}


\section{INTRODUCTION.}
\label{sec:intro}  % \label{} allows reference to this section
\subsection{Direct Observation of Exoplanets with a Space Telescope}

Since stars are many orders of magnitude brighter than planets, the lobes of the star's point spread function (PSF) will hide a planet even if the planet is resolvable by the optical system.  
% In order to directly image an exoplanet with a standard telescope the angular separation between the planet and the star must be very large (quote # here) and the planet must be very bright (quote mag here).
The lobes of the star's PSF must be suppressed in a way that still allows for the planet light to pass freely;  optical systems that accomplish this are referred to as coronagraphs\cite{krist_numerical_2016}.  There are many types of coronagraphs but most contain one mask in the focal plane to block the main stellar PSF and one mask in the pupil plane to deal with diffraction effects of the focal plane mask.  These masks create a `dark zone' (DZ) or `dark hole' (DH) in the focal plane of the coronagraph near the star where a planet can be observed.  In order to directly observe a rocky Earth-like planet, the starlight must be suppressed by a factor of $10^{10}$\cite{kasting_exoplanet_2009}.  For this level of suppression, deformable mirrors (DMs) are required to correct for the small manufacturing defects and misalignments in the optical system.  The magnitude of suppression is commonly referred to as the contrast which is calculated via
\begin{align}
    \text{contrast} = \frac{I_{coron}}{\max{(I_{direct})}}\label{fig:contrast}
\end{align}
where $I_{coron}$ is the intensity of the coronagraph image at the planet location and $I_{direct}$ is the peak intensity of the nominal image of the star without the focal plane mask (FPM) in place.  
% The High Contrast Imaging Testbed (HCIT) at the Jet Propulsion Laboratory operates in vacuum and has achieved contrast levels of $4\times 10^{-10}$\cite{seo_testbed_2019}.

Even with the coronagraph and deformable mirrors, large numbers of exposures are required to accumulate enough photons to characterize dim exoplanets.  The long observation times put strict stability requirements on the PSF.  In space, there are many sources of quasi-static wavefront error drifts that can degrade the PSF.  This paper attempts to inject and correct for space-like quasi-static drifts using the High-contrast imager for Complex Aperture Telescopes (HiCAT)\cite{soummer_high-contrast_2018} at the Space Telescope Science Institute (STScI).  HiCAT contains a classical Lyot coronagraph, two Boston Micromachine kilo (BMC) deformable mirrors, and an IrisAO PTT111L segmented aperture.  The monochromatic source is a Thorlabs MCLS1 diode laser at 638~nm and the broadband source is a Leukos SM-30-400 laser combined with a filter wheel.  The filter wheel has five 10~nm bandpass filters available at 610, 620, 640, 660, and 670~nm as well as a 6\% bandpass filter centered at 640~nm.  The simulator used to emulate the testbed (\texttt{catkit}\cite{noss_spacetelescopecatkit_2021}) is discussed in Fowler et. al. 2020\cite{fowler_generalized_2020} and Moriarty et. al. 2018 \cite{moriarty_high-contrast_2018}.  HiCAT can achieve contrast levels of $2.5\times10^{-8}$ ($6.5\times 10^{-8}$ 90\% of the time) using the monochromatic laser source and $6.3\times10^{-7}$ over a 20~nm band using the broadband laser source\cite{redmond_implementation_2021}. 
Here, we demonstrate maintenance of the dark zone contrast in the presence of a wavefront error drift at monochromatic levels of $2.5\times10^{-8}$ (IrisAO drift) and $5.9\times10^{-8}$ (all DMs drift), as well as $7.0\times10^{-7}$ at multiple wavelengths (BMC drift).  This work is an extension of Redmond et al. 2020\cite{redmond_implementation_2020} and Pogorelyuk et al. 2019\cite{pogorelyuk_dark_2019-1} which can be referred to for additional background and results. 

%  Often the planet can be dimmer than the static contrast obtained by the instrument so stability means stabilizing the image at levels that are a small fraction of the raw contrast so that one can calibrate that noise a posteriori.
% mention RST requirements?

\subsection{Quasi-static Optical System Drifts}
% \subsection{Quasi-static System Drifts}

The majority of telescopes performing direct imaging of exoplanets are ground-based and are thus limited by atmospheric seeing.  Atmospheric seeing creates fast, lower-order wavefront error (WFE) drifts that require DMs with high temporal frequencies.  Due to the seeing limit, ground based experiments are not as sensitive to non-common path (NCP) aberrations and often do not use the science detector for wavefront control\cite{sivaramakrishnan_sensing_2008}.  Now that we are introducing coronagraphs in the stratosphere\cite{mendillo_picture-c_2019} and in space\cite{noauthor_nancy_nodate}, we need to begin addressing the higher-order, quasi-static drifts that these experiments will experience during an observation period.  These slow drifts degrade the contrast\cite{soummer_speckle_2007} and can cause false planet identification\cite{spalding_its_2018}.  Space-based high contrast imagers are very sensitive to NCP aberrations so the first constraint is that we must use Focal Plane Wavefront Sensing and Control (FPWSC) to correct for the quasi-static drifts.  Note that this assumes there are DMs in the back-end optics which is not the case for the Hubble Space Telescope or the James Webb Space Telescope which depend on modelling and post-processing to remove drift effects\cite{debes_pushing_2019}$^,$\cite{perrin_updated_2018}.  

The two main space telescopes we will consider are the Roman Space Telescope (RST)\cite{krist_numerical_2016} and the proposed Large UV/Optical/IR Surveyor (LUVOIR)\cite{pueyo_luvoir_2019} mission.  RST has a 2.4~m monolithic glass primary mirror which has a very high thermal mass.  The current observing plan for the telescope involves slewing to a bright reference star to generate the dark zone and then to the dimmer science target to acquire science images until the dark zone degrades.  Any time on a reference star is lost science time so if the dark zone can be maintained using the science target, that is preferable.  The current observing scenario (Krist et al. 2020\cite{noauthor_roman_nodate}) also includes changes in the roll angle multiple times while on a science target.  This changes the thermal environment which can cause the wavefront to drift from its original state. LUVOIR will be a segmented aperture telescope which introduces a new set of issues as outlined in Coyle et al. 2019\cite{coyle_large_2019}.  Since each segment has its own mounting structure, thermal effects introduce modes with higher spatial frequencies than when compared to traditional monolithic primary mirrors.  In addition, each segment has three controllable degrees of freedom (DOF), piston/tip/tilt (PTT), and three un-controllable DOF (x, y, clocking).  The PTT actuators must be repeatable and stable (as well as the support structure) in order to avoid introducing WFE into the optical system.  Maintaining picometer-level stability across such a large aperture on long time-scales without any active drift correction will be extremely difficult.       

Dark zone maintenance (DZM) is still in the early stages of hardware testing so we manually inject drifts into the optical system that loosely represent the quasi-static drifts we expect to see on space telescopes such as RST and LUVOIR.  For the DM injecting the drift, each actuator performs a random walk.  The drifting DM can be either the BMC DMs or the IrisAO segmented aperture (or all three).  Chromatic effects of drifts can also cause issues so we study the DZM performance at multiple wavelengths.  We show that the DZM algorithm can potentially increase science observing time by eliminating the need of a reference star and relax stability requirements for large aperture segmented telescopes.   

Section \ref{sec:dhm} walks through the DZM algorithm and how it has been expanded to allow for different drifts and multiple wavelengths.  Next we cover the experimental results from relevant experiments on HiCAT in Sec.~\ref{sec:results}.  Lastly, Sec.~\ref{sec:conclusions} provides an overview of the DZM results so far and outlines the plan for future work.  

\section{Dark Zone Maintenance.}\label{sec:dhm}
% General method( explain dither, drift)
% Stuff unique to this paper:
% resume to choose starting dither
% 0 gain for first n iterations
% explain 'broadband' methodology

\subsection{Dark Zone Creation and Metrics}\label{subsec:dh_dig_metrics}
Before the dark zone can be maintained, a high contrast across the dark zone must be achieved.  This is commonly referred to as `digging a dark hole'.  HiCAT uses a stroke-minimization controller\cite{pueyo_optimal_2009}$^,$\cite{groff_methods_2015} and a pair-wise probe estimator\cite{groff_methods_2015} to initially generate the dark zone; this is discussed in detail in N'Diaye et al 2015 \cite{ndiaye_high-contrast_2015} and Soummer et al 2019 \cite{soummer_high-contrast_2019}.  Since HiCAT uses a classical lyot coronagraph, the control region is a 360$^{\circ}$ annulus.  For experiments in this paper, the control radius is from 5.8--9.8$~\lambda / D_{lyot}$ and the mean contrast is the spatially averaged contrast in the control region which is expressed as    
\begin{align}
    \text{mean contrast} &= \mu = \frac{1}{n}\sum_{k,l} \frac{I_{coron}(k,l)}{\max{(I_{direct})}}
\end{align}
where $n$ is the number of pixels in the dark zone and $k,l$ are the indices of the pixels in the dark zone.  As discussed in Redmond et al. 2020\cite{redmond_implementation_2020}, the drift and maintenance algorithm are not turned on until the pair-wise probe estimator and stroke-minimization controller (PPSM algorithm) generate the high-contrast region.  The final DM command and electric field estimate are used as the starting point for the dark zone maintenance algorithm.  The DZM algorithm is not designed to generate a dark zone and thus its performance is largely limited by the starting contrast provided by the PPSM algorithm.

\subsection{Dark Zone Maintenance Algorithm}\label{subsec:dhm}
% General method( explain dither, drift)
% Stuff unique to this paper:
% resume to choose starting dither
% 0 gain for first n iterations
% explain 'broadband' methodology
For a block diagram of the DZM algorithm as well as the initial algorithm derivation, see Redmond et al. 2020\cite{redmond_implementation_2020}.  Since the algorithm has been expanded to allow for different drifts and wavelengths, the derivation is covered again in this paper with some minor adjustments.  For this dynamical system, the state is the open loop drifting electric field in the dark zone ($x$), the measurement is the closed loop intensity at the science camera in the dark zone ($z$), and the control is the BMC DM command ($u$).  Note that the IrisAO is only used for drift injection and not control.  There are $n$ pixels in the dark zone and $m$ actuators are available for the control.  The electric field is a complex number so the real and imaginary components are split to facilitate working with real matrices.  The DZM algorithm uses an Extended Kalman Filter (EKF) to estimate the open loop electric field at the desired wavelengths and an Electric Field Conjugation (EFC) controller to determine the DM command to correct for the drift.  The EKF operates in a pixel-wise manner so the real and imaginary components for each pixel are stacked for both the state variable and the Jacobians.  This looks like
\begin{align}
    G_{i} &= \sqrt{t_{exp}}\begin{bmatrix}
        \Re\left(\frac{\partial x_0}{\partial u}\right)\\\\
        \Im\left(\frac{\partial x_0}{\partial u}\right) \\
        \vdots\\
        \Re\left(\frac{\partial x_n}{\partial u}\right)\\\\
        \Im\left(\frac{\partial x_n}{\partial u}\right)
    \end{bmatrix}\\\\
    x&= \begin{bmatrix}
        \Re\left(x_0\right)\\
        \Im\left(x_0\right) \\
        \vdots
    \end{bmatrix}
\end{align}
where 0--$n$ is the pixel in the dark zone, $i$ is the wavelength, and $G$ is the BMC Jacobian.  Note that $G$ the total BMC Jacobian where the Jacobians for BMC DM1 and BMC DM2 are stacked horizontally ($G = [G_{DM1}~~G_{DM2}]$).  The EKF estimator operates in units of counts so the Jacobians are in units of [sqrt(counts/s)/nm] and must be multiplied by the square-root of the exposure time ($\sqrt{t_{exp}}$) each iteration. 
%***************************************

For this paper, the drift is injected as a random walk of each DM actuator and can be expressed as
\begin{align}
    u^{k+1}_{drift} = u^{k}_{drift} + \mathcal{N}(0,\sigma _{drift} ^2 \mathcal{I}) \label{eq:drift_command}
\end{align}
where $k$ is the iteration, $\mathcal{I}$ is the identity matrix, and $\sigma_{drift}$ is the standard deviation of the normal distribution ($\mathcal{N}$).  For the BMC DMs, $\sigma_{drift}$ is a single number; for the IrisAO,  $\sigma_{drift}$ is broken up into $\sigma_{P}$, $\sigma_{T}$, $\sigma_{t}$ since each segment has piston ($P$), tip ($T$), and tilt ($t$) capabilities (PTT).  When using the IrisAO to inject drift, each degree of freedom drifts independently and are not constrained to be from the same distribution.

The nonlinear state space model is described as
\begin{align}
    x^{k+1} &= f(x^k) + w^k\\
    x^k_{CL} &= x^k + Gu^k \\
    y^{k+1} &= h(x^k,u^{k}) + n^k = x^k_{CL_{Re}}\circ x^k_{CL_{Re}}+ x^k_{CL_{Im}}\circ x^k_{CL_{Im}}+ n^k
\end{align}
where $\circ$ is the Hadamard operator, $u^k$ is the DM command for the current iteration, $x^k_{CL}$ is the closed loop electric field, and $w^k$, $n^k$ are the process and observation noise terms.  As with the BMC Jacobians, the DM command $u$ is a stacked vector containing the command for both BMC DMs ($u = [u_{DM1}~~u_{DM2}]^T$).  Since we are injecting drift using the DMs, the process noise term at each iteration is
\begin{align}
 w^k = G(u_{drift}^{k+1} - u_{drift}^{k}) + G_{P}(u_{P}^{k+1} - u_{P}^{k})+ G_{T}(u_{T}^{k+1} - u_{T}^{k}) + G_{t}(u_{t}^{k+1} - u_{t}^{k})
\end{align}
where $G_{PTt}$ are the IrisAO Jacobians for piston, tip, tilt and $u_{PTt}$ are the IrisAO piston, tip, tilt drift commands.  At each time step the nonlinear model is linearized to produce
\begin{align}
    \hat{\bar{x}}^k &= F^k x^{k-1} + w^k\\
    \hat{\bar{z}}^k &= H^k x^k_{CL} + G u^k+ n^k\\
    F^k &= \frac{\partial f}{\partial x^k}\Big|_{x^k =\hat{x}^{k-1}}\\
    H^k &= \frac{\partial h}{\partial x^k}\Big|_{x^k =\hat{x}^{k|k-1}}
\end{align}
where $H^k$ is the observation matrix and $F^k = \mathcal{I}$ 
is the state transition matrix.  The linearized estimate of the state and the measurement are shown by $\hat{\bar{x}}^k$ and $\hat{\bar{z}}^k$; they are a stepping stone to demonstrate how the observation matrix and state transition matrix are formulated and are not used in the EKF.  Since our state is the open loop electric field there are no dynamics in the system and the time dependant component is contained in the process noise term.   

The Extended Kalman Filter is a nonlinear estimator with memory solving a maximum likelihood estimation (MLE) problem. The EKF takes into account the previous time step biases in addition to the current estimate residual\cite{sun_efficient_2019}.  The first step in the EKF is to calculate the estimate and variance for the current time step given the estimate for the previous time step 
\begin{align}
    \hat{x}^{k|k-1} &= F^k \hat{x}^{k-1}  \\
    P^{k|k-1} &= F^k P^{k-1} (F^k)^T + Q^k 
\end{align}
where $Q^k$ is the process noise.  For the initial time step, we have the electric field estimate from the PPSM algorithm and initialize $P^{0-1} = 0$.  

The process noise must contain the total predicted drift in the system.  Due to the additive property of random matrices, the covariance of the sum is the sum of the covariances\cite{noauthor_expected_nodate}.  In the presence of a random walk drift of each BMC DM actuator, the BMC process noise is expressed as
\begin{align}
    Q^k_{BMC} = G G^T  \sigma_{drift}^2 % t_{exp}
\end{align}
as discussed in Pogorelyuk et al. 2020\cite{pogorelyuk_effects_2020}.  Here, $\sigma_{drift}$ is the standard deviation of the random drift command applied to each BMC actuator as shown in Eq.~\ref{eq:drift_command}.  The formulation is similar for the IrisAO except we must account for the three degrees of freedom,
\begin{align}
    Q^k_{iris} = \left(G_P G_P^T  \sigma_{P}^2 +  G_T G_T^T  \sigma_{T}^2  +  G_t G_t^T  \sigma_{t}^2 \right)%t_{exp}
    \label{eq:iris_cov}
\end{align} 
where $G_P$, $G_T$, and $G_t$ are the piston, tip, and tilt Jacobians for the IrisAO.  To get the total process covariance matrix we add the contributions from each DM to get
\begin{align}
    Q^k = Q^k_{BMC} + Q^k_{iris}.
\end{align}

Up next we calculate the Kalman gain which is used to determine the electric field estimate and covariance for the current iteration
\begin{align}
    K^k &= P^{k|k-1} (H^k)^T \left(H^k P^{k|k-1} (H^k)^T  + R^k\right)^{-1}\\
    \hat{x}^k &= \hat{x}^{k|k-1} + K^k \left[z^k - h(\hat{x}^{k|k-1},u^k)\right]\label{eq:xhat}\\
    P^k &= P^{k|k-1} - K^k H^k P^{k|k-1}
\end{align}

where $R^k$ is a square matrix containing the nonlinear measurement estimate of the state (open loop intensity) on the diagonal; $R^k = \text{diag}(\hat{y}^{k|k-1}) = \text{diag}(h(\hat{x}^{k|k-1},u^k))$.  Note that the EKF does not have access to the entire command applied to the BMC DMs, only the EFC command and dither: 
\begin{align}
    u^k &= \Delta u^k_{opt} + \delta u^k_{dither}\\
    \delta u^k_{dither} &= \mathcal{N}(0,\sigma_{dither}^2\mathcal{I})
\end{align}

where $\Delta u^k_{opt} $ is the optimal DM command determined by EFC and $\delta u^k_{dither} $ is the dither command.  Pogorelyuk et al. 2019 \cite{pogorelyuk_dark_2019} and Riggs et al. 2016 \cite{riggs_recursive_2016} discuss the risk of using an EKF without any probe images as it can cause the estimate to converge to the wrong value.  The dither command acts as a small probe which increases the phase diversity of the electric field between iterations to improve the estimate while not significantly degrading the contrast. Also, by using a random dither every image can still easily be used as a science image in post-processing.

The HiCAT hardware experiments take an open loop image every $k_{ol}$ iterations to track the open loop contrast and estimate error.  The open loop intensity estimate and BMC DM command are calculated as
\begin{align}
    \hat{I}^k_{OL} &= \hat{x}^k_{\Re}\circ \hat{x}^k_{\Re}+ \hat{x}^k_{\Im}\circ \hat{x}^k_{\Im}\\
    u_{OL}^{k} &= u_0 + u^k_{drift} \label{eq:u_open_loop}
\end{align}
where $\hat{x}$ is calculated in Eq.~\ref{eq:xhat} and $u_0$ is the initial DM command obtained from the PPSM algorithm as discussed in Sec.~\ref{subsec:dh_dig_metrics}.  Note that due to the way the state vector is assembled, $\hat{x}^k_{\Re} = \hat{x}^k[0,2,\hdots, 2n-2]$ and $\hat{x}^k_{\Im} = \hat{x}^k[1,3,\hdots, 2n-1]$.

Once the estimate of the open loop electric field is obtained, EFC\cite{giveon_broadband_2007} is used to determine the optimal DM command to minimize the closed loop field present via
\begin{align}
    \Delta u^{k+1}_{opt} &= -\beta \left(G^TG + \alpha \mathcal{I}\right)G^T \hat{x}^k
\end{align}
where $\alpha$ is the Tikhonov regularization parameter to avoid excessively large DM commands and $\beta$ is the control gain.  The EKF operates by storing information from the previous estimates and needs to accumulate a certain amount of data before the estimate will converge\cite{redmond_implementation_2020}.  The control gain ($\beta$) is set to zero for the first 10--20 iterations, depending on the speed of the drift, to allow the estimator to converge prior to applying an EFC command.  An important note is that the total control command update\cite{pogorelyuk_dark_2019} changes slightly when the open loop electric field assumed to drift and used as the state variable.  In a traditional EFC use case\cite{sun_efficient_2019} where the closed loop electric field is the state variable and the open loop electric field is assumed to be constant, the DM command us updated via $u^{k+1}_{control} += \Delta u^{k+1}_{opt}$.  In DZM, once the EFC correction is obtained the DM command applied to the BMC mirrors follows as
\begin{align}
    u^{k+1}_{control} &= u_0 + \Delta u^{k+1}_{opt} + \delta u^{k+1}_{dither}\\
    u^{k+1}_{tot} &= u^{k+1}_{control} + u^{k+1}_{drift}
\end{align}
where $u_0$ is the DM command obtained from the final iterations of PPSM, $u^{k+1}_{control}$ is the total closed loop DM command, and $u^{k+1}_{tot}$ is the total command sent to the DMs including the drift.

To perform DZM at multiple wavelengths only a minor adjustment to the algorithm is required.  For each iteration, an image is taken at each of the $l$ wavelengths of interest using the narrowband filters described in Sec.~\ref{sec:intro}.  The same DM command is used for all wavelengths.  The EKF is run on each wavelength individually producing an estimate of the electric field at each wavelength ($\hat{x}^k(\lambda_i)$).  Note that the Jacobians ($G$), observation matrix ($H^k$), and covariance matrices ($Q^k,P^k,R^k$) are now all wavelength dependant.  An EFC command is then calculated for each wavelength using the EKF output and the average command is used for control as shown by
\begin{align}
    \Delta u^{k+1}_{opt} &= \frac{1}{l}\sum_{i=1}^l  u^{k+1}_{opt}(\lambda_i)
\end{align}
where $u^{k+1}_{opt}(\lambda_i)$ is the EFC output for $\lambda_i$.  Note that if desired, weights can be applied to prioritize certain wavelengths.

\section{Laboratory Results.}\label{sec:results}

\subsection{General Comments}
\subsubsection{Drift metrics}
On the HiCAT testbed, the majority of the iteration time is due to file writing and thus the drift per iteration is a better metric when extrapolating to other experiments than the drift per minute.  Both values will be provided for each experiment as the HiCAT environment is changing in real time.  Each iteration there are 2 required images taken, one image for the estimator and one image after the EFC correction, as well as two optional images, an open loop image and a direct image without the FPM.  The image after the EFC correction is not strictly necessary and could be removed to shorten iteration time.  In addition to the drift rate, the final DM surface root mean square (RMS) and peak-to-valley (PV) are provided for the drifting DM. 

\subsubsection{BMC DM error}\label{sec:bmc_ol_error}
The DZM performance is degraded when open loop images are taken during the experiments.  This is due to the large change in the command applied to the BMC DMs for the open loop image vs. the closed loop image.  We have seen that the BMC DMs do not return to the exact shape when an intermediate command is applied.  Since we are dealing with very high contrasts, this effect begins to matter.  In a real observing scenario, open loop images would not be collected so this effect should not be an issue for any mission.         

% other PTT drift case had pinhole in so a lot more exposures were required
\subsubsection{Pinhole}
In July, a pinhole was temporarily placed after the laser launch to try and filter some of the fiber effects seen when using the broadband laser source.  This reduced the throughput by a factor of five thus requiring longer exposure times.  It was removed shortly thereafter due to how long experiments were taking.  In this paper, the pinhole was in place for Sec.~\ref{subsec:iris_ptt} results but removed for Sec.~\ref{subsec:bb_drift} and \ref{subsec:all_dms_drift} results. 

\subsubsection{Dither calibration}\label{sec:dither_cal}
In the case where only the BMC DMs are drifting, it is relatively easy to determine the drift-dither relationship\cite{redmond_implementation_2020}.  When the IrisAO drift is added, this relationship becomes harder to characterize.  For the results presented here, the dither is tuned using the smallest drift that produces a noticeable change in the open loop electric field in the span of 30~mins.  An initial dither is selected as the value that produces the final contrast from the PPSM algorithm via Fig.~8 in Redmond et al. 2020\cite{redmond_implementation_2020}.  If the estimator performs poorly, the dither is increased until it is stable.  From there, if the drift is changed the dither is scaled with it.  Due to the model matching in Redmond et al. 2020\cite{redmond_implementation_2020}, the dither can often be tuned in simulation only requiring minor adjustments when we transfer to the testbed.  This method of dither tuning proved effective but further investigation into a more efficient approach is under way.       

\subsection{BMC Drift at Multiple Wavelengths}\label{subsec:bb_drift}
To build on previous BMC drift results, we performed a similar experiment with the BMC DMs performing a random walk drift at multiple wavelengths.  Note that the HiCAT testbed has not been optimized for broadband performance and is limited to a contrast of $3\times10^{-7}$ even when monochromatic FPWC is performed using the broadband source and a single narrowband filter.  Due to limitations when using the broadband laser source we have not yet attempted true broadband DZM.  For this experiment we used the 640~nm and 660~nm filters which results in a 20~nm band spanning 30~nm.  Two exposures are averaged for each image resulting in a total exposure time of 0.1~s and a total iteration time of 30~s.  

Since we are operating at $10^{-7}$ for this experiment, larger dither and drift values of $\sigma_{dither} = $300~pm/iter and $\sigma_{drift}=$~30~pm/iter are used respectively.  When the iteration time is accounted for, the drift rate scales to  $\sigma_{drift}=$~59.2~pm/min.  The control gain $\beta$ is zero for the first 10 iterations and then set to one.  There is no spike in contrast at the beginning of the experiment so 10 iterations is long enough for the estimator to converge and short enough so that the open loop electric field does not significantly drift in that time.  The final BMC drift command for each DM has an RMS of 0.93~nm and a PV of 6.15~nm.

\begin{figure}[H]
	\centering
 	\begin{subfigure}[b]{0.45\textwidth}
 	    \centering
    	\includegraphics[scale=0.37]{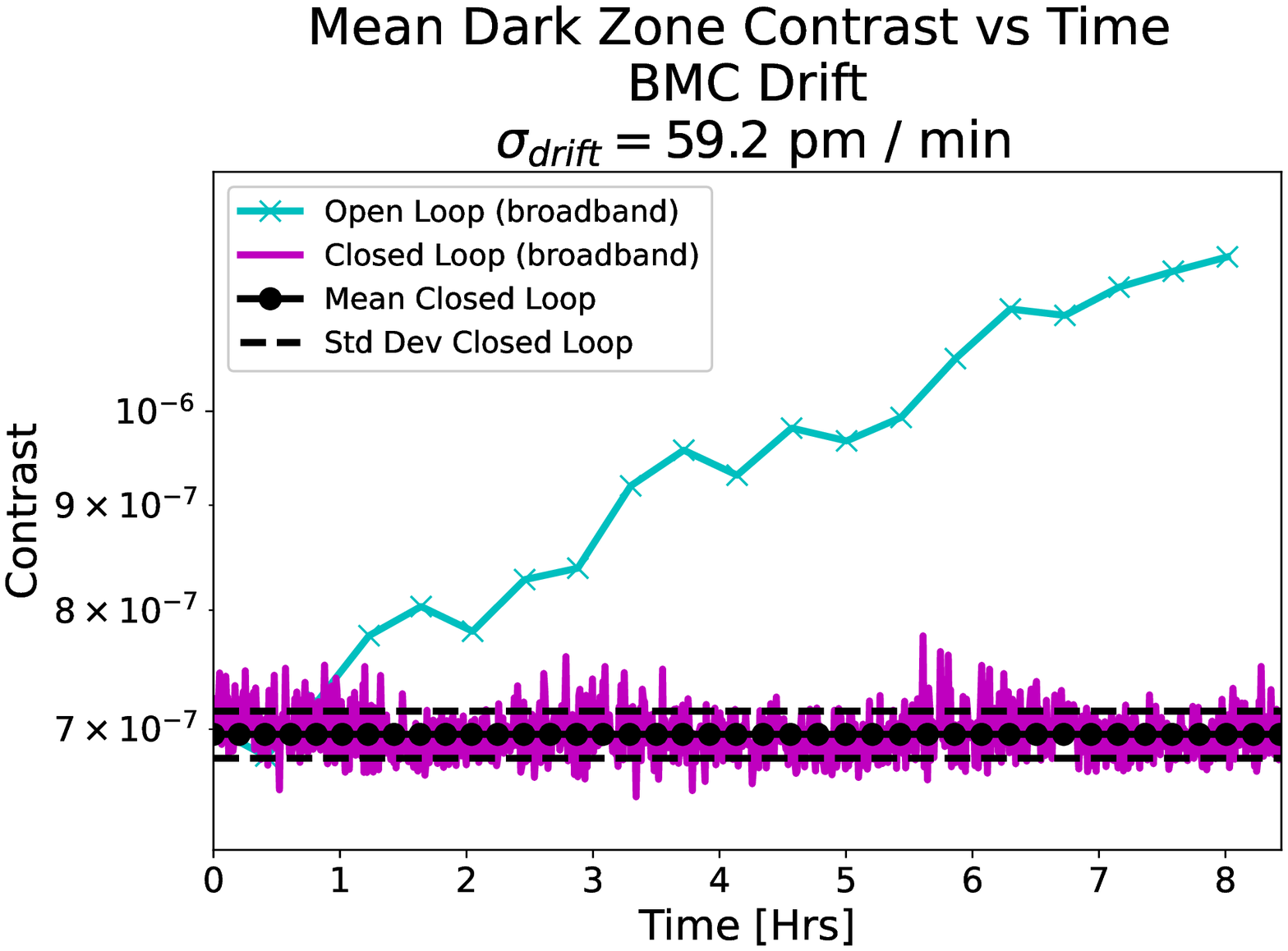}
    	\captionsetup{justification=centering}
	    \caption{Broadband contrast vs time}
	    \label{fig:bb_cont_vt}
 	\end{subfigure}
 	\begin{subfigure}[b]{0.1\textwidth}
 	\centering
 	\end{subfigure}
 	\begin{subfigure}[b]{0.5\textwidth}
 		\centering
    	\includegraphics[scale=0.34]{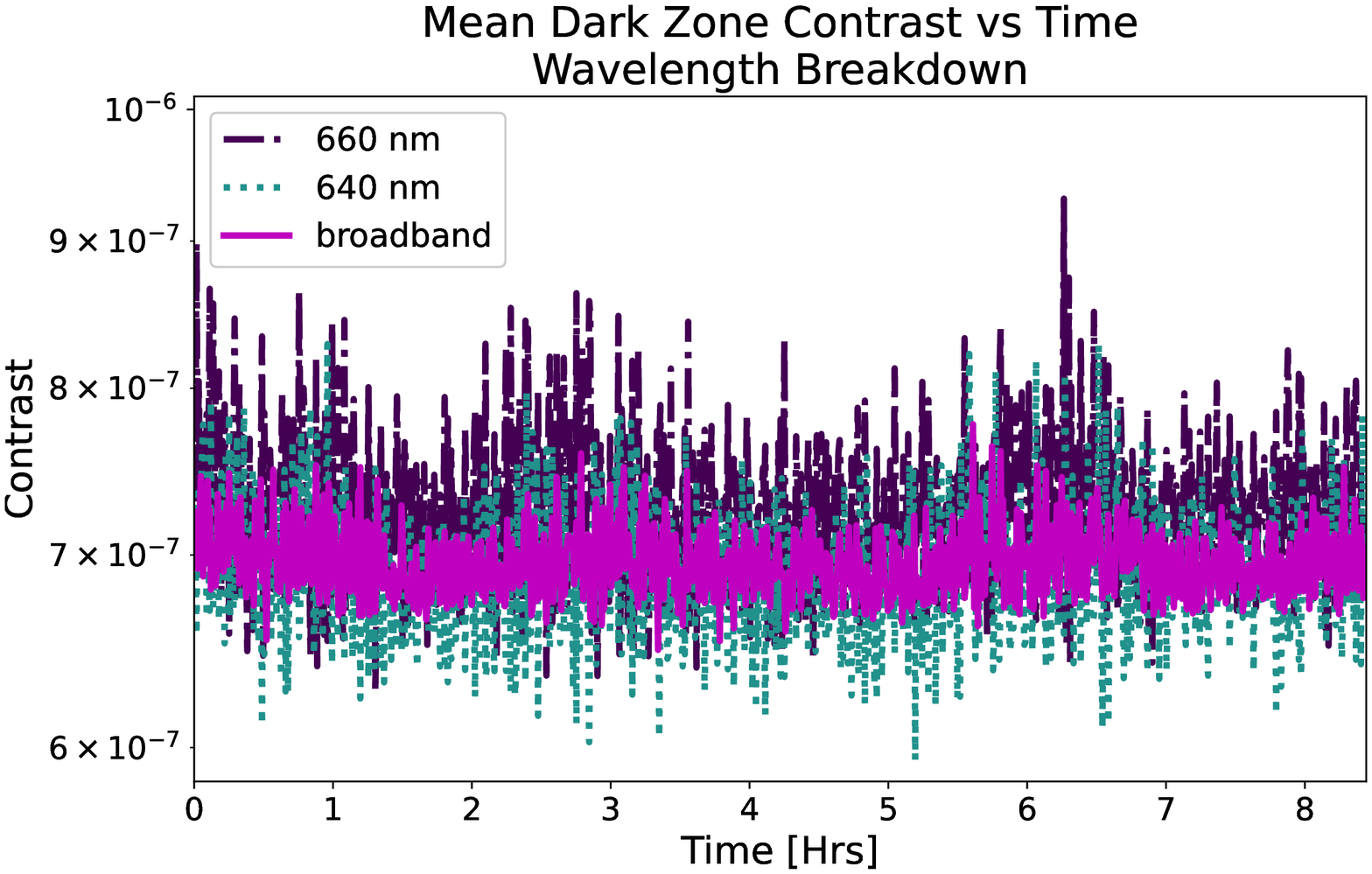}
    	\captionsetup{justification=centering}
	    \caption{Wavelength breakdown contrast vs time.}
	    \label{fig:lam_cont_vt}
 	\end{subfigure}
 	\caption{Contrast vs time plots for multi-wavelength maintenance in the presence of BMC random walk drift.  The left plot (\ref{fig:bb_cont_vt}) shows the open and closed loop broadband contrasts, with the cyan x's and magenta curves respectively, for the duration of the experiment.  The mean closed loop broadband contrast is maintained at $7.0\times10^{-7}$ (black dots) with a standard deviation of $1.8\times10^{-8}$ (black dashes).  The open loop broadband contrast degrades by a factor of 1.72 to $1.2\times10^{-6}$.  The right plot (\ref{fig:lam_cont_vt}) shows the closed loop contrast vs time for the wavelengths controlled.  The 660~nm contrast (mauve dashes) is slightly worse than the 640~nm contrast (teal dots) but they are within 5\% of each other.  For reference, the broadband contrast is provided in \ref{fig:lam_cont_vt} in magenta.}
 	\label{fig:bb_maint}
 \end{figure}
Figure \ref{fig:bb_maint} shows the contrast vs. time plots for multi-wavelength dark zone maintenance where the BMC DMs are drifting.  As shown in Fig.~\ref{fig:bb_cont_vt}, a mean broadband dark zone contrast of $7.0\times10^{-7}$ is maintained for 8.5~hrs with a standard deviation of $1.8\times10^{-8}$.  This is one of the best results we have seen so far in that the standard deviation of the mean dark zone contrast is only 2\% of the mean dark zone contrast.  Certainly, the fact that we are operating at $10^{-7}$ is part of what makes this possible as the other drifts in the system such as temperature, humidity, and air turbulence have less of an effect at this level.  The open loop broadband contrast increases by a factor of 1.72 to $1.19\times10^{-6}$ by the end of the experiment. Figure \ref{fig:lam_cont_vt} shows the wavelength breakdown for the closed loop mean dark zone contrast during the experiment. Here we can see that the 640~nm contrast is slightly better than the 660~nm contrast but they are within 5\% of each other.    

\begin{figure} [htb]
	\centering
   \includegraphics[scale=0.43]{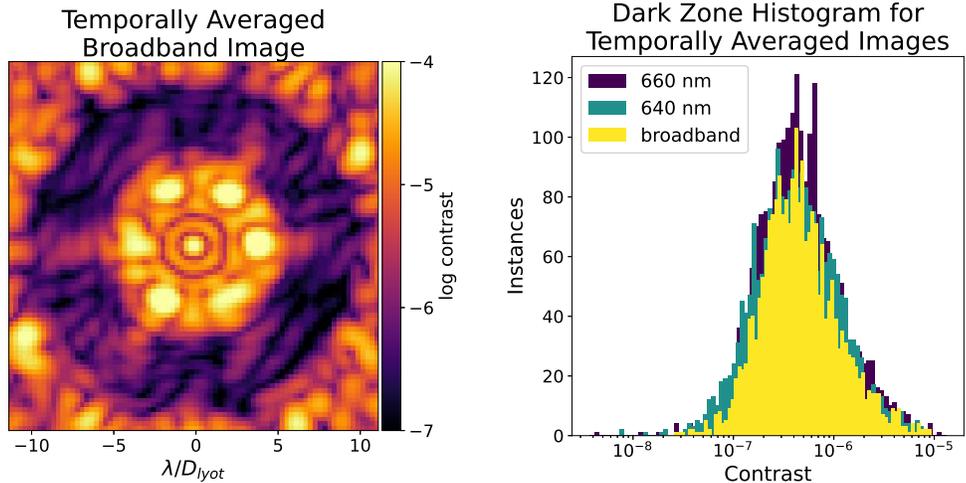}
   \caption[Multi-wavelength maintenance histogram] 
   { \label{fig:bb_hist}Temporally averaged image distributions.  The left panel shows the temporally averaged broadband image for all 1000 iterations where $D_{lyot}$ is the diameter of the Lyot stop.  There is a distinct residual of the six-fold spoke pattern induced by the IrisAO which is the main source of the lower contrast speckles and positively skewed distributions. This spoke pattern residual is more common when the broadband laser source is used and work is ongoing to remove it.  The right panel shows the histograms for the 660~nm, 640~nm, and broadband temporally averaged images.  Note that this histogram has a log scale on the x-axis.  All histograms are similar showing that all wavelengths are being equally maintained in the broadband DZM algorithm.  }
\end{figure}

In addition to the spatially averaged contrast data, we decided to look at the contrast distribution in the dark zone.  For an individual iteration, the distribution has a large positive skew making it difficult to provide a quality metric for each iteration.  When we averaged over all 1000 images from the experiment, the distribution becomes less skewed as shown in Fig.~\ref{fig:bb_hist}.  The left panel in Fig.~\ref{fig:bb_hist} shows the temporally averaged broadband image.  There is a clear residual of the six-fold spoke pattern created by the IrisAO; this defect is mainly seen when the broadband laser source is used and is a driver of the broadband contrast.  In the right panel of Fig.~\ref{fig:bb_hist} we provide the histograms for the temporally averaged images for each wavelength as well as the combined broadband case.  The characteristics of these histograms are provided in Table \ref{tab:bb_stats} which shows an equal level of DZM across the band.

\begin{table}[htb]
\caption{Temporally averaged image statistics.} 
\label{tab:bb_stats}
\begin{center}       
\begin{tabular}{|c|c|c|c|} %% this creates two columns
\rule[-1ex]{0pt}{3.5ex} Wavelength & Mean Contrast in DZ & Std Dev of Contrast in DZ & Skew of Contrast Distribution \\
\hline
660~nm &$7.2\times10^{-7}$ &$9.9\times10^{-7}$ & 4.62 \\  
640~nm &$6.9\times10^{-7}$& $9.1\times10^{-7}$ & 4.11 \\  
broadband &$7.1\times10^{-7}$ & $9.1\times10^{-7}$ & 4.13 
\end{tabular}%\label{tab:bb_stats}
\end{center}
\end{table} 

% std dev is only 2% of mean

% no pinhole

% want contrast for each lambda
\subsection{IrisAO Piston Tip Tilt Drift}\label{subsec:iris_ptt}
% include humidity plot?
The IrisAO PTT drift experiment is targeted for future generation segmented telescopes such as LUVOIR\cite{pueyo_luvoir_2019} but can also loosely represent primary mirror drifts on RST\cite{noauthor_nancy_nodate}.  This is a monochromatic experiment where the piston, tip, and tilt of each IrisAO segment drift independently.  For this experiment, the drift in each DOF has approximately the same weight with $\sigma_P=$~5~pm/iter and $\sigma_{T,t}=$~5~nrad/iter.  The dither used is $\sigma_{dither} = 100$~pm/iter and was chosen as outlined in Sec.~\ref{sec:dither_cal}.
A pinhole was temporarily placed after the laser launch to try and filter some of the fiber effects.  This reduced the throughput by a factor of five thus requiring larger exposure times.  To achieve the desired signal-to-noise (SNR), each image consists of 16 exposures with the total exposure time being 0.48~s.  A relatively high SNR was chosen for this experiment so the number of exposures could be drastically reduced.  The iteration time is $\sim30$~s creating a drift rate of $\sigma_P=$~9.5~pm/min and $\sigma_{T,t}=$~9.5~nrad/min.  The final IrisAO drift command has a surface RMS of 0.19~nm and a surface PV of 1.06~nm. 

\begin{figure} [htb]
	\centering
   \includegraphics[scale=0.45]{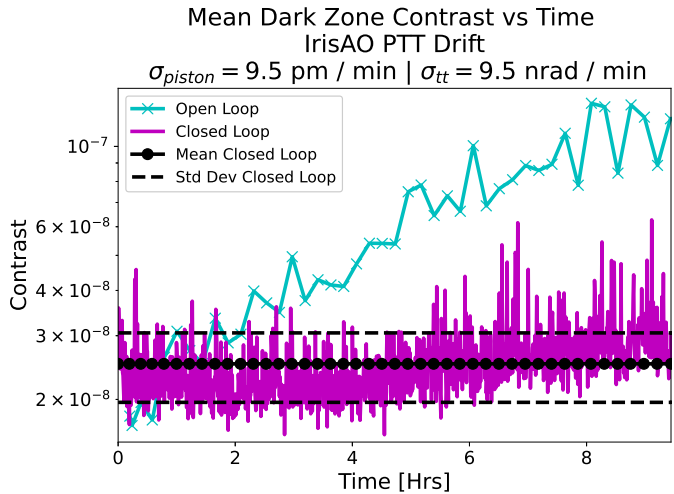}
   \caption[IrisAO PTT Drift] 
   { \label{fig:iris_ptt} IrisAO PTT drift. The mean contrast of the dark zone is shown in magenta with the mean and standard deviation of the red curve shown by the dashed and dotted black lines respectively.  The cyan xs mark the mean contrast in the dark zone for the open loop DM command.  The open loop mean contrast degrades by a factor of 4.75 while the closed loop mean contrast remains at the initial value of $2.5\times10^{-8}$ within a standard deviation of $5.5\times10^{-9}$.}
\end{figure}

Figure~\ref{fig:iris_ptt} shows the mean dark zone contrast evolution during the experiment. The experiment runs overnight for 9.5~hrs maintaining a closed loop contrast of $2.5\times10^{-8}$ (black dots) within a standard deviation of $5.5\times10^{-9}$ (black dashes).  The open loop contrast drifts by a factor of 4.75 to $1.2\times10^{-7}$ (cyan x's) during the experiment.  The control gain $\beta$ is set to zero for the first 20 iterations and then set to one for the duration of the experiment.  As shown by the first 10~mins of the plot in Fig.~\ref{fig:iris_ptt}, there is no large overshoot as the estimator converges and the closed loop electric field does not significantly drift while the controller is off.  The closed loop contrast (magenta curve in Fig.~\ref{fig:iris_ptt}) is extremely stable for the first 4.5 hours and then begins to deviate slightly as the open loop contrast becomes significantly larger.  This small deviation is due to an overly-cautious Tikhonov parameter ($\alpha$) when determining the EFC correction combined with the BMC DM error effect discussed in Sec.~\ref{sec:bmc_ol_error}.  If we split the experiment into two blocks (0--4.5~hrs and 4.5--9.5~hrs) we see that the mean contrast increases by 20\% from $2.3\times10^{-8}$ to $2.7\times10^{-8}$ after 4.5~hrs and the standard deviation increases by 54\% from $3.8\times10^{-9}$ to $5.9\times10^{-9}$.  While there is a distinct increase in both the mean contrast and standard deviation, we are not concerned as: (1) we are aware of the cause and can increase $\alpha$ / decrease frequency of open loop images for future experiments and (2) the absolute increase is still very small and manageable.    

\FloatBarrier

\begin{figure} [htb]
	\centering
   \includegraphics[scale=0.4]{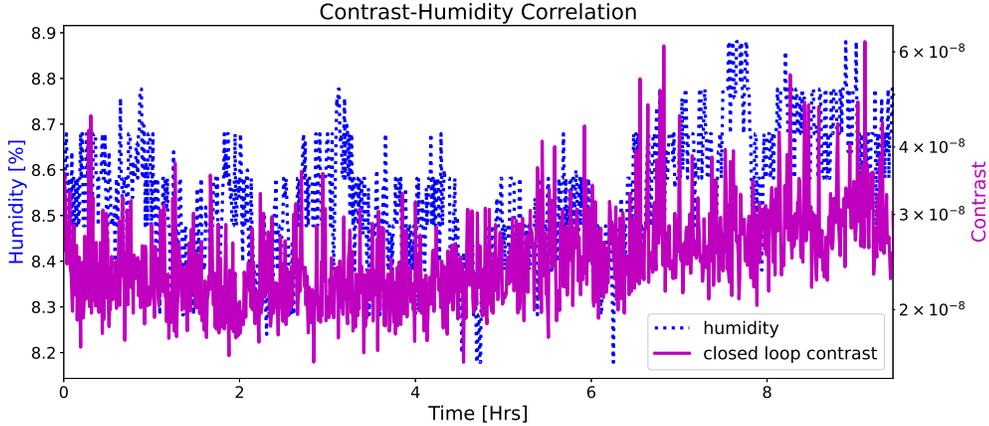}
   \caption[IrisAO drift contrast-humidity correlation] 
   { \label{fig:humidity} Contrast-humidity correlation.  The blue dotted line shows the humidity in the HiCAT enclosure during the experiment.  The mean dark zone contrast shown in magenta starts trending up when the humidity increases at the 4.5~hr mark. }
\end{figure}

HiCAT also tracks the temperature and humidity during experiments.  The WFE is known to be quite sensitive to changes in humidity.  As shown in Fig.~\ref{fig:humidity} the mean DZ contrast seems to track the humidity.  The humidity drift is not captured by the process noise covariance matrix thus the EKF will not try to correct for it.  The contrast-humidity correlation could be a coincidence and the driving factor may still be the controller tuning issues but further investigation is warranted.    

\subsection{BMC and IrisAO Drift}\label{subsec:all_dms_drift}

\begin{figure} [htb]
	\centering
   \includegraphics[scale=0.45]{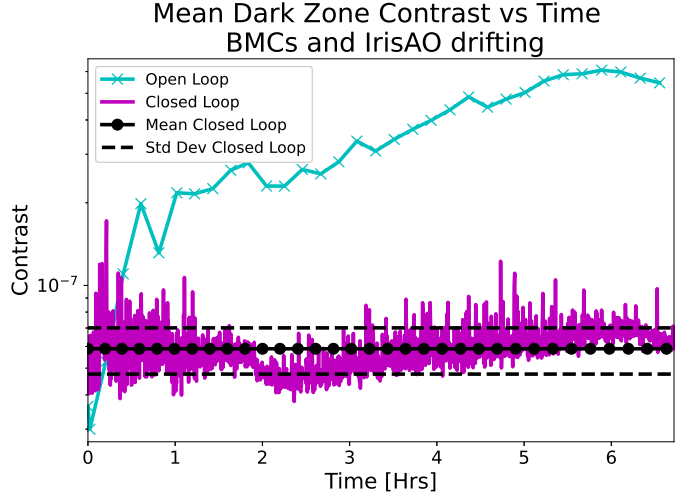}
   \caption[BMCs and IrisAO Drift] 
   {\label{fig:all_drift} BMCs and IrisAO PTT drift. The mean contrast of the dark zone is shown in magenta with the mean and standard deviation of the magenta curve shown by the dotted and dashed black lines respectively.  The cyan x's mark the mean contrast in the dark zone for the open loop DM command.  The open loop mean contrast degrades by a factor of 9.25 while the closed loop mean contrast remains at the initial value of $5.9\times10^{-8}$ within a standard deviation of $1.1\times10^{-8}$.}
\end{figure}

This experiment aims to study the effect of two different simultaneous drifts.  In this experiment both BMC DMs are drifting with $\sigma_{drift} = 10$~pm/iter as well as the IrisAO with $\sigma_{P} = 0.8$~pm/iter and $\sigma_{Tt} = 0.8$~nrad/iter.  The dither used is $\sigma_{dither} = 150$~pm/iter and since the drift was predicted to be relatively fast, the control gain $\beta$ was only set to zero for the first five iterations.  This was likely too few iterations as we see a small spike in the closed loop contrast in Fig.~\ref{fig:all_drift} when the controller turns on while the estimator is still converging.  Unlike in Sec.~\ref{subsec:iris_ptt}, the number of exposures is capped at three for this experiment.  With three exposures per image, the total exposure time is 0.024~s and the total iteration time is 15~s.  It should be noted that the pinhole is removed for this experiment which aides in reducing the iteration time.  At 15~s per iteration, the drift rates can be expressed as $\sigma_{drift} = 39.4$~pm/min for the BMCs and $\sigma_{P} = 3.1$~pm/min, $\sigma_{Tt} = 3.1$~nrad/min for the IrisAO.  

The final BMC drift command for each DM has an RMS of 0.40~nm and a PV of 2.86~nm.  For the IrisAO, final drift command has an RMS of 0.04~nm and a PV of 0.28~nm.  The final command is near the resolution of the IrisAO and in future, larger IrisAO drifts should be used so that the covariance matrices more accurately represent the drift.  Even though we are near the resolution of the IrisAO, we can confidently say it is drifting by comparing these results to those from Redmond et al. 2020, Fig.~7\cite{redmond_implementation_2020}. For the 2020 experiment, only the BMCs are drifting ($\sigma_{drift} = 10$~pm/iter) and the initial contrast is approximately the same (1.5$\times$ larger than the combined experiment). The open loop contrast reaches $1.4\times10^{-7}$ in 1000 iterations for the BMC only case but reaches a contrast of $4.4\times10^{-7}$ in 1000 iterations when the IrisAO drift is added in (Fig.~\ref{fig:all_drift} at 4.14~hrs).  This difference is large enough to conclude that the IrisAO is introducing WFE into the system, even if it is not exactly what we are predicting.  

As shown by the contrast vs time plot in Fig.~\ref{fig:all_drift}, the open loop contrast drifts by a factor of 9.25 reaching $5.5\times10^{-7}$ after 1500~iterations or 7~hrs. The spatially averaged closed loop contrast has a mean of $5.88\times10^{-8}$ and a standard deviation of $1.1\times10^{-8}$ over the entire experiment.  This performance is slightly worse than the experiment in Sec.~\ref{subsec:iris_ptt} for a number of reasons.  When applying the final DM command from the PPSM algorithm, HiCAT reliably resumes around $6\times10^{-8}$ but sometimes it is better or worse.  In Sec.~\ref{subsec:iris_ptt} we start from a very good resume at $2\times10^{-8}$ and thus the mean contrast is better.  Also, while the standard deviation is at the $10^{-8}$ level when all DMs are drifting, it is still only 20\% of the mean which is on par with the results from Sec.~\ref{subsec:iris_ptt}.  As in Sec.~\ref{subsec:iris_ptt}, the Tikhonov parameter is too conservative for this experiment as well which allows some of the drift to leak through.  We are also still taking open loop images relatively frequently which degrades the performance.  Another note is that since the we are near the resolution of IrisAO, its behaviour is not accurately depicted by the process noise covariance matrix (Eq.~\eqref{eq:iris_cov}) which may add to the drift leakage.  

%15s iter
 
\section{Conclusions and Future Work.}\label{sec:conclusions}

Future space missions will need a reliable method for dealing with quasi-static high-order WFE drifts.  In this paper we demonstrate the DZM capabilities in a variety of situations.  In order to do spectroscopy on directly observed exoplanets, the contrast must be consistent over a wide wavelength range.  In Sec.~\ref{subsec:bb_drift} we use DZM to correct for BMC random walk drift at a rate of $\sigma_{drift}=30$~pm/iter and maintain the contrast of a 20~nm band at $7.0\times10^{-7}$ for 8.5~hrs.  We also provide the wavelength-dependant statistics for the temporally averaged dark zone demonstrating that the closed-loop contrast is uniformly maintained.  Large segmented aperture telescopes introduce new challenges that HiCAT aims to address.  Using the IrisAO, we demonstrate the ability to maintain a contrast of $2.5\times10^{-8}$ in the presence of a PTT random walk drift with a rate of $\sigma_P=5$~pm/iter and $\sigma_{Tt}=5$~nrad/iter.  Space telescopes will have multiple drift sources with different temporal and spatial frequencies.  To represent this effect, Sec.~\ref{subsec:all_dms_drift} shows the results for an experiment where the BMC DMs drift at a rate of $\sigma_{drift}=10$~pm/iter and the IrisAO drifts at a rate of $\sigma_P=0.8$~pm/iter, $\sigma_{Tt}=0.8$~nrad/iter.  The closed loop contrast is maintained at $5.9\times10^{-8}$.

In the Sections \ref{subsec:iris_ptt} and \ref{subsec:all_dms_drift} we see the drift leaking into the closed loop electric field because the EFC Tikhonov parameter is too large.  At the $10^{-8}$ level we also see a performance degradation due to the open loop images which require a large change in the BMC DM command. Section \ref{subsec:iris_ptt} also shows a potential sensitivity to sub-percent changes in the humidity when at $2.5\times10^{-8}$.  Future work will address these issues by spending more time tuning the controller, reducing the number of open loop images taken, and performing a humidity sensitivity study at a range of contrasts.  In addition, we plan on determining the maximum allowable IrisAO drift and relating that to the LUVOIR stability requirements.  The HiCAT broadband source will be upgraded in late 2021 at which point we will perform broadband DZM on a wider wavelength band.  For the broadband DZM, we also plan on investigating the broadband EFC method outlined in Give'on et al. 2007\cite{giveon_broadband_2007} to replace the average EFC method currently used.      

% future experiment: turn off humidity control to calibrate effect?

%  tryoptimal dither technique (rather than the random dither used in this paper) described in Sun 2019 \cite{sun_efficient_2019}.

\acknowledgments % equivalent to \section*{ACKNOWLEDGMENTS}       
 
This work was supported in part by the National Aeronautics and Space Administration under Grant \\
80NSSC19K0120 issued through the Strategic Astrophysics Technology/Technology Demonstration for Exoplanet Missions Program (SAT-TDEM; PI: R. Soummer).

E.H.P. is supported by the NASA Hubble Fellowship grant \#HST-HF2-51467.001-A awarded by the Space Telescope Science Institute, which is operated by the Association of Universities for Research in Astronomy, Incorporated, under NASA contract NAS5-26555.

% References
\bibliography{zotero_references} % bibliography data in report.bib
\bibliographystyle{spiebib} % makes bibtex use spiebib.bst

\end{document}